# Thermopower enhancement by fractional layer control in 2D oxide superlattices

Woo Seok Choi[1,2,3,*], Hiromichi Ohta[4], and Ho Nyung Lee[1,**]

[1]Materials Science and Technology Division, Oak Ridge National Laboratory, Oak Ridge, TN 37831, USA

[2]Department of Physics, Sungkyunkwan University, Suwon, Gyeonggi-do 440-746, Korea

[3]IBS Center for Integrated Nanostructure Physics, Institute of Basic Science (IBS), Sungkyunkwan University, Suwon, Gyeonggi-do 440-746, Korea

[4]Research Institute for Electronic Science, Hokkaido University, N20W10, Sapporo 001-0020, Japan

[*]e-mail: choiws@skku.edu.

[**]e-mail: hnlee@ornl.gov.

We have investigated two-dimensional thermoelectric properties in transition metal oxide heterostructures. In particular, we adopted an unprecedented approach to direct tuning of the 2D carrier density using fractionally $\delta$-doped oxide superlattices. By artificially controlling the carrier density in the 2D electron gas that emerges at a $La_xSr_{1-x}TiO_3$ $\delta$-doped layer, we demonstrate that a thermopower as large as 408 $\mu$V K$^{-1}$ can be reached. This approach also yielded a power factor of the 2D carriers 117 $\mu$Wcm$^{-1}$K$^{-2}$, which is one of the largest reported values from transition metal oxide based materials. The promising result can be attributed to the anisotropic band structure in the 2D system, indicating that $\delta$-doped oxide superlattices can be a good candidate for advanced thermoelectrics.

Thermoelectric phenomenon is indispensable in understanding the transport nature of itinerant charge carriers and their interaction with the crystal lattice. In practical sense, it is utilized to convert heat into electric power (Seebeck effect) or to generate temperature gradient from electricity (Peltier effect). Therefore, thermoelectric power generation is considered as one of the most important technologies for sustainable energy. The thermoelectric efficiency is often quantified by the thermoelectric figure of merit $ZT = (S^2\sigma/\kappa)T$, where $S$, $\sigma$, $\kappa$, and $T$ are the Seebeck coefficient or thermopower, the electrical conductivity, the thermal conductivity, and the absolute temperature, respectively. In order to achieve a high $ZT$ value, the prerequisites of materials are large $S$ and $\sigma$, and low $\kappa$. However, phenomenologically, $S$ and $\log \sigma$ are inversely proportional to each other, which poses difficulty in maximizing the $ZT$ value.[1] Moreover, $\sigma$ and $\kappa$ are linearly proportional to each other if the thermal transport is dominated by charge carriers instead of phonons, again limiting the controllability of $ZT$. Therefore, in most cases, achieving a delicate balance among $S$, $\sigma$, and $\kappa$ is the key to achieving the maximum $ZT$ value.

Until recently, the majority of the research on thermoelectric materials has been devoted to the conventional semiconductors such as $Bi_2Te_3$ and PbTe, partially owing to their high efficiency.[1] However, their volatile nature especially at high temperatures and the use of toxic elements (such as Pb, Sb, Se, and Te) prevented these materials from being a ubiquitous choice. On the other hand, recent studies reported that transition metal oxides (TMOs) can be a promising candidate for highly efficient thermoelectrics with high $|S|$ and $\sigma$, and hence, the high power factor ($PF = S^2\sigma$).[2, 3] More importantly, in most cases, the strong correlation uniquely found in TMOs can be a useful tool for developing new thermoelectric materials. For example, $Na_xCoO_2$ showed a large $|S|$ due to a strong electronic correlation,[4] and low dimensional Nb-doped $SrTiO_3$ exhibited an unexpectedly large $PF$[5-7] by a dimensional crossover of the polaronic behavior.[8, 9] More recently, it has also been shown that TMO superlattices can have minimized thermal conductivity due to the crossover from incoherent to coherent phonon scattering, suggesting that it can be beneficial for thermoelectric applications.[10] Owing to the versatile physical properties of the TMOs, the thermoelectric properties can be further investigated in terms of complex electronic and crystallographic structures, such as disproportionate band structure, effective mass anisotropy, dimensionality, and formation of heterointerfaces.[2, 3] We further emphasize that the additional degree of freedom in TMOs (i.e., charge, spin, orbital, and lattice) and strong coupling among them could lead to an unprecedented thermoelectric behavior with novel physical origins.[11]

In addition to the selection of appropriate materials system, the geometry of samples also plays a critical role in determining the thermoelectric efficiency.[12, 13] In particular, a large |$S$| could be achieved due to the modified electronic structure near the Fermi level ($E_F$) in low dimensional structures. For example, quantum confinement effect on the thermoelectric property was observed in a PbTe/Pb$_{1-x}$Eu$_x$Te multiple quantum well structure,[12] and enhancement of |$S$| due to the electron filtering effect was observed in InGaAs based superlattices.[14] More recently, SrTiO$_3$ based 2D structures also exhibited an interesting low-dimensional effect.[5, 6] As recent technical advances in synthesizing oxide heterostructures provide an unprecedented opportunity for realizing various low dimensional structures with the atomic precision, we can use such structures as a test bed for enhancing the thermoelectric properties.

In order to develop highly efficient, low dimensional TMO thermoelectrics, we focus on the strongly correlated 2D perovskite oxide materials. In particular, previous studies found the interface between the Mott insulator LaTiO$_3$ (LTO) and the band insulator SrTiO$_3$ (STO) bears a two-dimensional electron gas (2DEG). The 2D system exhibited intriguing transport and optical properties including electronic reconstruction,[15, 16] anomalous $T$-dependent metallic behavior,[17] multi-carrier/multi-channel conduction,[18-20] 2D superconductivity and quantum critical behavior.[21-23] More interestingly, it has been recently shown that the carrier density could be effectively controlled by a selective band filling through controlled doping of the 2DEG layer.[20]

In this paper, we present carrier-density-tuned 2D thermoelectric phenomena for a direct comparison with the 3D case for a wide range of carrier densities. As shown in Fig. 1, this could be achieved by fabricating fractional superlattices (SLs) using pulsed laser epitaxy (PLE). We carefully modified the carrier concentration in the LSTO ($\delta$-doped) layer by incorporating different concentrations of $La^{3+}$ ions within the layer. Using the deliberate crystal design, we could achieve a precise control on the carrier density ($n$) of LSTO as was previously achieved in the 3D case.[24] Furthermore, we observed a large enhancement of $|S|$ due to the 2D confinement effect, for the same $n$, without sacrificing much of the electron mobility in TMO SLs. The enhancement of $|S|$ resulted in a 300% enhancement in power factor for a moderate value of $n$, suggesting that the low dimensional effect together with the controllability of $n$ is an important factor to be considered in oxide thermoelectric materials.

Fractional $\delta$-doping method was used to fabricate $[(La_xSr_{1-x}TiO_3)_1/(SrTiO_3)_{10}]_{10}$ (LSTO/STO, $x$ = 0.25, 0.50, 0.75, and 1.00) SLs. Note that the SLs were adopted to amplify the effect from each 2DEG layer. The number of 2DEG layers in each sample was fixed to 10. Details on the PLE-based controlled $\delta$-doping method can be found elsewhere.[20] To measure the transport properties of the SLs, we made a direct Ohmic contact with indium to the metallic 2D LSTO layers in the van der Pauw geometry using ultrasonic soldering. Temperature- ($T$-) dependent $S$ was measured by conventional steady-state method ($\Delta T$ ~2 K) using a cryogenic refrigerator. Other transport properties were characterized using a physical property measurement system (Quantum Design Inc.).

The transport properties of the SLs are shown in Fig. 2 as a function of $x$. The value of $S$ at room $T$ was extracted from linear fitting of the $\Delta V$-$\Delta T$ curves as shown in the lower left inset of Fig. 2a. For all the samples, $S$ revealed a negative value indicating $n$-type conduction. The values of $|S|$ were comparable to those reported from Nb:STO/STO SLs.[5, 6] With increasing La$^{3+}$ incorporation within the $\delta$-doped layer ($x$), $|S|$ gradually decreased. The $T$-dependent $S$ curves shown in the upper right inset indicate that $|S|$ decreases with decreasing $T$. In slightly doped bulk LSTO ($x \leq 0.1$), $T^2$-dependence was observed for $|S|$-$T$ curves which could be later attributed to the enhancement of the electron-phonon coupling with the decrease in $n$.[25] For our SLs, however, we could not observe any characteristic $T$-dependent behavior (other than the typical linear $T$-dependence observed for diffusion thermoelectricity) that could be attributed to unconventional physical mechanism. The room $T$ carrier density ($n$) and mobility ($\mu$) values of the SLs (corresponding to the 2D layers) are shown in Fig. 2b. Here, $n$ is normalized by the nominal total thickness of the 2D layers as discussed below. The linear increase of $n$ with $x$ is clearly observed as expected, while $\mu$ does not show any considerable $x$-dependence.

Conventionally, the $ZT$ value is calculated based on bulk (3D) $\sigma$, and thus, an effective thickness of the 2D layer is required to calculate $n$ in cm$^{-3}$. In this way, the thermoelectric properties of low dimensional materials could be properly understood. While it is practically impossible to experimentally obtain the exact thickness of the 2D layer in LSTO/STO SLs,[5, 6] one may consider the well-studied transport property of the SLs in order to estimate the thickness. LTO/STO and LSTO/STO interfacial systems at low $T$ (< ~50K) have to be understood in terms of multi-carrier or multi-channel transport.[18-20] According to magnetic field dependent Hall resistance measurements, the so-called "low-density-high-mobility carriers" located in STO, *i.e.*, away from the interface, become active at low *T*. This is due to enhanced dielectric screening of carriers assisted by the largely enhanced dielectric constant of STO especially at low *T*.[26] On the other hand, at room *T*, "high-density-low-mobility carriers" located at the interface (or the LSTO layer) play a dominant role in the transport property. The dielectric constant of STO is an order of magnitude smaller at room *T* than that at low *T*, suggesting a minimized spill-over effect of conduction electrons into the STO sides. In addition, oxygen vacancies have been carefully removed throughout the SLs, and we believe that the STO spacing layers remain insulating. Therefore, to understand the room *T* carrier dynamics of the *δ*-doped SLs, we have considered only the high-density-low-mobility carriers confined at the LSTO layer, with the nominal thickness of 1 u.c. (~4 Å).

The linear dependence of *n* with *x* was observed in 3D LSTO,[24, 25] where each $La^{3+}$ ion ideally adds one electron to the system by modifying $Ti^{4+}$ ($d^0$) into $Ti^{3+}$ ($d^1$) ($0 < x \leq 0.95$). Note that if all $Sr^{2+}$ ions are substituted by $La^{3+}$ ($LaTiO_3$), the bulk sample becomes a Mott insulator, although one would expect $1.68 \times 10^{22}$ $cm^{-3}$ of carrier density from simple reasoning. For the 2D *δ*-doped layer, such trend in bulk, *i.e.*, linear increase of *n* with *x*, was similarly observed. However, the SL does not become insulating even in the case of *x* = 1, possibly due to the smearing out of a small number of carriers even at room *T*. The measured *n* of our LTO/STO SL (corresponding to the 2D layers) is $8.1 \times 10^{21}$ $cm^{-3}$, suggesting that only half of the electrons compared to the bulk contribute to the actual conduction in 2D. Note that such decrease in *n* can also be attributed to the surface/interface depletion of the carriers.[27]

The decrease in $x$ (or $n$) yields an increase in $|S|$ in 2D. This behavior coincides with the phenomenological behavior for 3D thermoelectric materials, suggesting that this phenomenological model can be universally applied for the lower dimensional cases as well. Note that such a direct comparison study could only be realized due to the fabrication of fractional SLs. For $x = 0.25$ SL, $|S|$ increased up to 408 $\mu$V K$^{-1}$, which is ~2.5 times larger compared to $x = 1.00$ SL, indicating that tuning $n$ in 2D system is an effective method to enhance the $|S|$ value.

For more detailed analyses, we directly compared the transport property of LSTO for 3D and 2D (filled blue circles) cases, as summarized in Fig. 3 and Fig. 4. The data for 3D bulk LSTO (empty squares) for systematic $n$ values is taken from a previous study by Okuda *et al.*[25] First, we note that the electrical conductivity is rather lower in our SLs than that in the bulk LSTO (Fig. 3a) for the same $n$. This is mainly due to the reduced $\mu$ which could be attributed to the reduced dimensional or confinement effect. On the other hand, $\mu$ has a completely different tendency for bulk and SL samples, as shown in Fig. 3b. For the 3D case, the decrease of $\mu$ with decreasing $n$ was observed and attributed to the tendency of localization of the carriers, possibly due to the enhanced electron-phonon coupling at lower $n$ as previously discussed.[25] Similar decrease in $\mu$ with decreasing $n$ at room $T$ has also been observed in another study on La doped STO bulk.[28] Note that the values of $\mu$ at room $T$ were similar for other electron dopants in STO such as Nb or oxygen vacancies,[5, 6] although the decreasing trend with decreasing $n$ was less evident.[29] In our 2D LSTO case, however, we did not find any distinct $n$-dependent behavior, which suggests that such an $n$-dependent localization effect does not prevail for a very wide range of $n$. Nevertheless, the dimensional confinement induced localization seems to affect the carrier transport for all the SLs. Indeed, even scattering with La$^{3+}$ ions does not seem to play a large role at room $T$, as $\mu$ is almost constant and independent of $n$ (or $x$). This might be attributed to the fact that for a $\delta$-doped layer, electron-phonon coupling does not depend on $n$ anymore, due to the atomic confinement of 2D carriers.[10]

The most important observation regarding the thermoelectric effect in the 2D oxide SLs is the enhancement of $|S|$ due to the reduced dimensionality. We first note that the qualitative trend, *i.e.*, the increase of $|S|$ with the decrease of $n$, is the same for both 3D and 2D. While a phenomenological trend has been also observed in heavily Nb-doped STO,[2] our direct and deliberate control of $n$ through fractional layer structuring reported here for the first time can open a door to novel TMO thermoelectric heterostructures.

More interestingly, $|S|$ of our SLs is much larger than that of the bulk samples for the same $n$ values (Fig. 4a). In particular, more persistent $|S|$ is observed as the $n$ is increased over a wide range, as compared to the bulk samples. To estimate the enhancement by the dimensional crossover over a wide range of $n$ (which was not accessible in bulk), we adopted the curve (grey line), which was used to theoretically explain the $n$-dependence of $|S|$ in heavily Nb-doped STO bulk.[2] Note that this curve was a modification of the Jonker relationship and explains the 3D data by Okuda *et al.* quite well especially for low $n$.[30] Based on this estimation of $|S|$ in 3D, $S_{2D}/S_{3D}$ was calculated (empty diamond symbols in red). $S_{2D}/S_{3D}$ shows a drastic increase for $x = 0.50$ compared to $x = 0.25$ SLs. Above $x = 0.50$, it more or less saturates. However, it should be noted that the estimation of $|S|$ in 3D overestimates the last few data points by Okuda *et al.*, suggesting that $S_{2D}/S_{3D}$ can be even much larger than what shown here, for higher $n$ values ($x = 0.75$ and 1.00 SLs). In addition to the orbital degeneracy of the Ti $3d$-$t_{2g}$ band and strong correlation attributed to the large $|S|$ in oxide single crystals, the reduced dimensionality enforces disproportionate (anisotropic) band structure, which further enhances $|S|$. The structural anisotropy becomes larger as $n$ increases, which should most probably lead to an increased anisotropy of the carrier transport (*e.g.*, the in-plane conductivity linearly increases with increasing $x$, while there is no linear relationship for the out-of-plane conductivity),[18] and the deviation of the |*S*| values between 2D and 3D also increases with increasing $n$. Such a large enhancement of |*S*| for large $n$ values increases the *PF* of the SLs substantially as shown in Fig. 4b. Compared to the largest value for 3D LSTO (35 $\mu$Wcm$^{-1}$K$^{-2}$), the *PF* value corresponding to the 2D layers reaches up to 117 $\mu$Wcm$^{-1}$K$^{-2}$ for $x = 0.75$ SL, which is more than a 300% enhancement over the values reported from the bulk samples. Moreover, this *PF* value is larger than that for most of the oxide heterostructures reported up to date (shown in triangles in Fig. 4b),[8, 28, 31-33] which can be mostly explained by 3D thermoelectric calculations,[2] especially at high $n$. This is mainly due to the 2D confinement effect together with a wide $n$ control. We again emphasize that such a large *PF* value could be achieved uniquely through systematic fabrication of $\delta$-doped SLs with controlled doping. Finally, since the thermal conductivity of the 2D layer could not be measured, we estimated it from the bulk value of 10 W K$^{-1}$m$^{-1}$ at room $T$,[2, 25] and obtained *ZT* values corresponding to the 2D layers ranging from 0.08 ($x = 0.25$) to 0.35 ($x = 0.75$).

In summary, we have shown that a highly efficient thermoelectric material can be designed by superlattice approach, in which the delicate balance between carrier density and thermopower is controlled by fractional control of 2D carriers. The general trend in the thermoelectric property of 3D transition metal oxides is still valid for the 2D SLs, and the maximum *S* value can be achieved when $n$ is at the lowest limit of conduction. However, |*S*| clearly increases in 2D compared to 3D, and one can achieve much larger *PF* values with 2D superlattices. The results demonstrated here indicate that modifying the dimension of carrier conduction is a way to tuning the thermoelectric properties of oxide materials beyond what the bulk counterpart can perform.

## Acknowledgements

We appreciate valuable discussion with J. H. Han and V. R. Cooper. This work was supported by the U.S. Department of Energy, Basic Energy Sciences, Materials Sciences and Engineering Division. WSC was supported by Samsung Research Fund, Sungkyunkwan University, 2013. HO is supported by JSPS-KAKENHI (25246023, 25106007).

Figure legends

Figure 1 | Schematics of fractionally-doped transition metal oxide superlattices. [(La$_x$Sr$_{1-x}$TiO$_3$)$_1$/(SrTiO$_3$)$_{10}$]$_{10}$ superlattice samples with controlled chemical compositions have been fabricated to detail the carrier conduction for controlling thermopower. As the portion of La in the 2DEG layer increases, both 2D carrier density and conductivity also increase, while 2D thermopower decreases.

Figure 2 | Transport properties of fractional superlattices. a, $S$ as a function of $x$. As $x$ decreases, $|S|$ increases, reaching over 400 $\mu$V K$^{-1}$ for $x = 0.25$. The lower left inset shows the potential difference as a function of $T$ gradient at 300 K. The upper right inset shows the $T$-dependence of $|S|$, which increases with increasing $T$. b, $n$ and $\mu$ as a function of $x$ at 300 K. $n$ has been normalized to the total effective thickness of the 2D layers. $n$ is linearly proportional to $x$, while $\mu$ does not exhibit any systematic change.

Figure 3 | Electronic transport behaviors of 2D and 3D thermoelectric oxides. a, $\sigma$ and b, $\mu$ as a function of $n$ for bulk samples (empty squares) and fractional SLs (filled blue circles). The reduced $\sigma$ is mainly due to the reduced $\mu$.

Figure 4 | Thermoelectric properties of 2D and 3D thermoelectric oxides. a, $S$ and b, $PF$ as a function of $n$ for bulk samples (empty squares) and fractional SLs (filled circles). While the general $n$-dependence is maintained, the fractional superlattices show highly enhanced thermoelectric properties, mainly due to the increased anisotropy. The thick grey line in a represents results from a theoretical calculation of the 3D data based on Jonker relationship.[2, 30] Right red axis in a is data for $S_{2D}/S_{3D}$, which are represented with the empty diamonds in red.

Various *PF* values for bulk (green triangles),[28] film (orange and cyan triangles),[31, 32] and superlattices (purple triangles)[8] are shown for comparison in **b**. The thick grey line represents data from a theoretical calculation for 3D.[2]

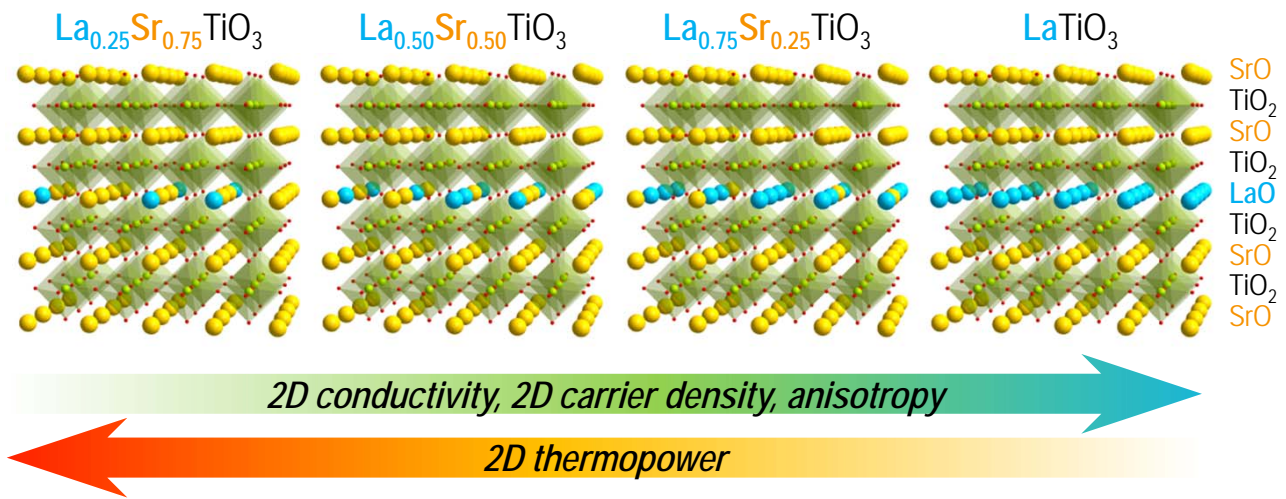

Figure 1. Choi *et al.*

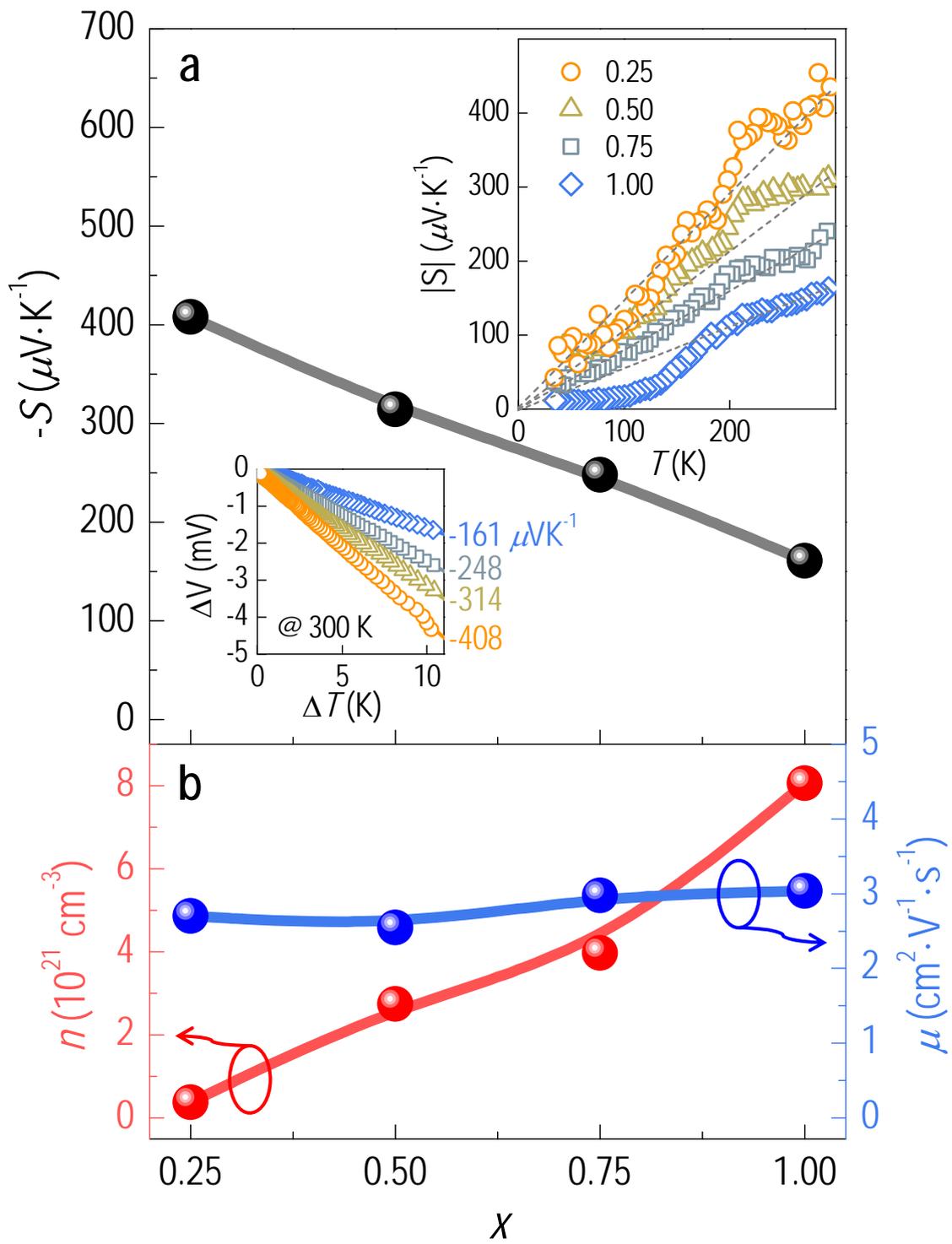

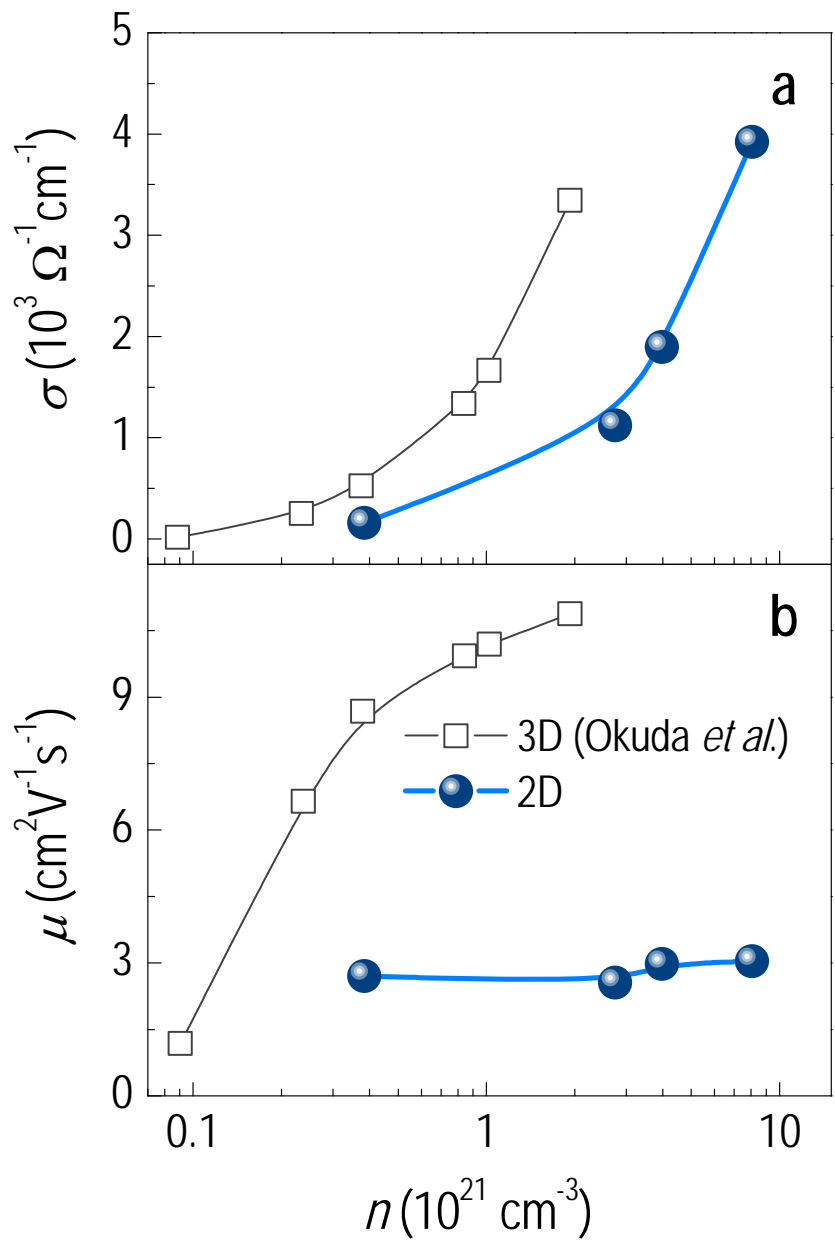

Figure 3.
Choi *et al.*

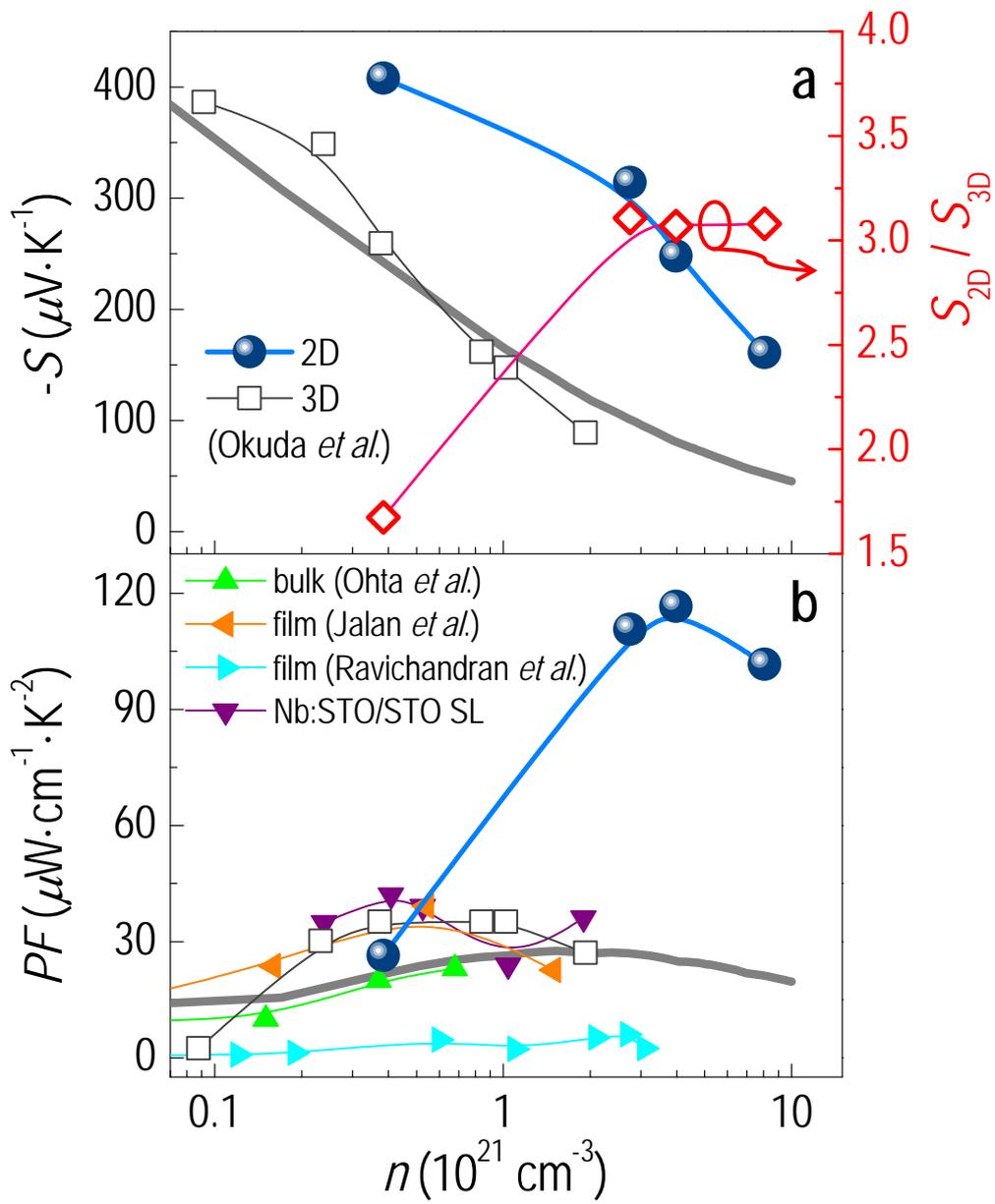

Figure 4. Choi *et al.*